\documentclass[prl,twocolumn,showpacs,superscriptaddress]{revtex4}%

\usepackage{graphicx}
\usepackage{dcolumn}
\usepackage{bm}

\begin{document}

\title{Precision determination of the $d\pi\leftrightarrow NN$ transition strength at threshold}

\author{Th.\,Strauch} 
\affiliation{Institut f\"{u}r Kernphysik, Forschungszentrum J\"{u}lich GmbH, D-52425 J\"{u}lich, Germany}
\author{F.\,D.\,Amaro}
\affiliation{Dept. of Physics, Coimbra University, P-3000 Coimbra, Portugal}%
\author{D.\,F.\,Anagnostopoulos} 
\affiliation{Dept. of Materials Science and Engineering, University of Ioannina, GR-45110 Ioannina, Greece}
\author{P.\,B\"uhler}
\affiliation{Stefan Meyer Institut for Subatomic Physics, Austrian Academy of Sciences, A-1090 Vienna, Austria}
\author{D.\,S.\,Covita}
\altaffiliation[present address: ]{I3N, Dept. of Physics, Aveiro University, P-3810 Aveiro, Portugal}%
\affiliation{Dept. of Physics, Coimbra University, P-3000 Coimbra, Portugal}%
\affiliation{Laboratory for Particle Physics, Paul Scherrer Institut (PSI), CH 5232-Villigen, Switzerland}
\author{H.\,Gorke}
\affiliation{Zentralinstitut f\"{u}r Elektronik, Forschungszentrum J\"{u}lich GmbH, D-52425 J\"{u}lich, Germany}
\author{D.\,Gotta} 
\altaffiliation[ ]{corresponding author: d.gotta@fz-juelich.de}%
\affiliation{Institut f\"{u}r Kernphysik, Forschungszentrum J\"{u}lich GmbH, D-52425 J\"{u}lich, Germany}
\author{A.\,Gruber}
\affiliation{Stefan Meyer Institut for Subatomic Physics, Austrian Academy of Sciences, A-1090 Vienna, Austria}
\author{A.\,Hirtl}
\altaffiliation[present address: ]{Universit\"atsklinik f\"ur Nuklearmedizin, Medizinische Universit\"at Wien, 
 1090 Vienna, Austria}
\affiliation{Stefan Meyer Institut for Subatomic Physics, Austrian Academy of Sciences, A-1090 Vienna, Austria}
\author{P.\,Indelicato}
\affiliation{Lab.\,Kastler\,Brossel\,(LKB),\,ENS,\,CNRS,\,UPMC-Paris\,6,\,Case\,74,\,4\,place Jussieu,\,F-75005 Paris,\,France}
\author{E.-O.\,Le\,Bigot}
\affiliation{Lab.\,Kastler\,Brossel\,(LKB),\,ENS,\,CNRS,\,UPMC-Paris\,6,\,Case\,74,\,4\,place Jussieu,\,F-75005 Paris,\,France}
\author{M.\,Nekipelov}
\affiliation{Institut f\"{u}r Kernphysik, Forschungszentrum J\"{u}lich GmbH, D-52425 J\"{u}lich, Germany}
\author{J.\,M.\,F.\,dos\,Santos}
\affiliation{Dept. of Physics, Coimbra University, P-3000 Coimbra, Portugal}%
\author{S.\,Schlesser}
\affiliation{Lab.\,Kastler\,Brossel\,(LKB),\,ENS,\,CNRS,\,UPMC-Paris\,6,\,Case\,74,\,4\,place Jussieu,\,F-75005 Paris,\,France}
\author{Ph.\,Schmid}
\affiliation{Stefan Meyer Institut for Subatomic Physics, Austrian Academy of Sciences, A-1090 Vienna, Austria}
\author{L.\,M.\,Simons}
\affiliation{Laboratory for Particle Physics, Paul Scherrer Institut (PSI), CH 5232-Villigen, Switzerland}
\author{M.\,Trassinelli}
\altaffiliation[present address: ]{Inst. des NanoSciences de Paris, CNRS UMR7588 and UMPC-Paris 6, F-75015 Paris, France}%
\affiliation{Lab.\,Kastler\,Brossel\,(LKB),\,ENS,\,CNRS,\,UPMC-Paris\,6,\,Case\,74,\,4\,place Jussieu,\,F-75005 Paris,\,France}
\author{J.\,F.\,C.\,A.\,Veloso}
\affiliation{I3N, Dept. of Physics, Aveiro University, P-3810 Aveiro, Portugal}%
\author{J.\,Zmeskal}
\affiliation{Stefan Meyer Institut for Subatomic Physics, Austrian Academy of Sciences, A-1090 Vienna, Austria}

\date{\today}

\begin{abstract}
An unusual but effective way to determine at threshold the $d\pi\leftrightarrow NN$ transition strength $\alpha$ 
is to exploit the hadronic ground-state broadening $\Gamma_{1s}$ in pionic deuterium, accessible by x-ray 
spectroscopy. The broadening is dominated by the true absorption channel $d\pi^-\rightarrow nn$, which is related 
to s-wave pion production $pp\rightarrow\,d\pi^+$ by charge symmetry and detailed balance. Using the exotic atom 
circumvents the problem of Coulomb corrections to the cross section as necessary in the production experiments. 
Our dedicated measurement finds $\Gamma_{1s}=(1171{+\,23\atop-\,49})$\,meV yielding $\alpha=(252{+\,5\atop -11})\,\mu$b. 
\end{abstract} 
\maketitle

Meson production and absorption at low energies plays a key role in developing methods within the 
framework of effective field theories such as chiral perturbation theory ($\chi PT$)\,\cite{Ber08}.  
Directly at threshold, experimental access to $NN\leftrightarrow NN\pi$ processes is provided both via 
the hadronic ground-state broadening $\Gamma_{1s}$ in pionic deuterium ($\pi$D) and pion production in 
nucleon-nucleon collisions. 

Considering only pure hadronic cross sections (denoted by $\tilde{\sigma}$\,), i.\,e., with the Coulomb interaction 
switched off to circumvent the divergence problem at threshold but with the particles keeping their physical mass, 
the production cross section is parametrised by\,\cite{Ros54}
\begin{eqnarray}
\tilde{\sigma}_{pp\rightarrow~\pi^{+}d}=\alpha\eta + \beta\eta^{3} + ...~\label{eq:pid_pp}
\end{eqnarray}
with $\eta =p^{*}_{\pi}c/m_{\pi}c^2$ being the reduced pion momentum in the $\pi d$ rest frame. 
For $\eta\rightarrow 0$ higher partial waves ($\beta,\,...$) vanish, and only the threshold parameter 
$\alpha$ contributes owing to pure s-wave production. Directly related is the reaction 
$np\rightarrow~\pi^{0}d$ because in the limit of charge independence the relation  
$2\cdot\sigma_{np\rightarrow~\pi^{0}d}=\sigma_{pp\rightarrow~\pi^{+}d}$ holds. 

\begin{figure}[t]
\begin{center} 
\resizebox{0.4\textwidth}{!}{\includegraphics{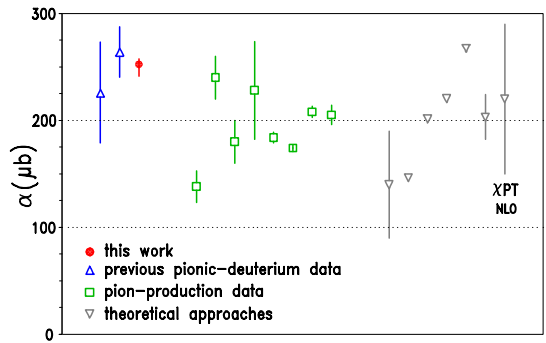}}
\caption{(color online). Threshold parameter $\alpha$. Points are shown in the same order as listed in 
          Table\,\ref{table:alpha}.}

\label{figure:alpha}
\end{center} 
\end{figure}

Values for $\alpha$ derived from pion-production\,\cite{Cra55,Ros67,Ric70,Aeb76,Hut91,Hei96,Dro98} and 
absorption experiments\,\cite{Rit91} scatter widely even when comparing recent data (Tab.\,\ref{table:alpha} 
and Fig.\,\ref{figure:alpha}). However, sometimes only statistical errors 
are given. The fluctuations suggest systematic uncertainties of about 10\% possibly stemming from 
normalization. In particular, the Coulomb corrections, mandatory to obtain the pure hadronic cross section, 
are a significant source of uncertainty\,\cite{Ros67,Rei69,Hut91}. 

As discussed for example by Lensky et al.\,\cite{Len06,Len07}, phenomenological 
descriptions\,\cite{Ros54,Kol66,Rei69,Afn74,Hor93,Nis96} may suffer from an incomplete knowledge of the 
contributing mechanisms, which in principle is avoided within the $\chi PT$ approach if enough terms are 
considered in the expansion. A recent calculation up to next-to-leading order (NLO) terms yields 
$\alpha^{NLO}=220\,\mu$b\,\cite{Len06} (Fig.\,\ref{figure:alpha}) and thus 
$\Im\,a_{\pi D}=5.65\cdot 10^{-3}\,m^{-1}_{\pi}$\,\cite{Len07} for the imaginary part of the $\pi$D 
scattering length. The uncertainty of about $\pm$\,30\% is expected to decrease to below $\pm$\,10\% 
by next-to-next-to-leading order (NNLO) calculations\,\cite{Bar09}. 

A measurement of $\Gamma_{1s}$ in pionic deuterium is equivalent to the determination of $\Im \,a_{\pi D}$\cite{Des54} 
being predominantly attributed to true pion absorption $\pi^-d\rightarrow nn$. Pion absorption at rest on the 
isospin I=0 nucleon-nucleon pair of the deuteron induces the transition 
$^{3}S_{1}[^{\,3}D_{1}]$(I=0)\,$\rightarrow$\,$^{3}P_{1}$(I=1), the inverse of which accounts for s-wave pion 
production in $pp\rightarrow d\pi^+$\,\cite{Bru51}. Therefore, $\Gamma_{1s}$ is a measure of the 
s-wave pion-production strength. 

In contrast to production experiments, the extraction of the threshold parameter $\alpha$ or $\Im\,a_{\pi D}$ 
from $\Gamma_{1s}$ avoids the problem of Coulomb corrections to the measured cross sections. However, previous 
x-ray experiments\,\cite{Cha9597,Hau98} are of limited statistics, and insufficient knowledge on the experimental 
resolution and cascade-induced broadening prevents a precise extraction of the pure hadronic width $\Gamma_{1s}$. 
Hence, a remeasurement of $\Gamma_{1s}$ was performed\,\cite{PSI98} aiming at an accuracy of at least the one 
expected from the forthcoming $\chi PT$ calculations. 

The complex pion-deuteron scattering length $a_{\pi D}$ is related to the 1s-state shift $\epsilon_{1s}$ 
and width $\Gamma_{1s}$ in $\pi$D by   
\begin{equation}
\epsilon_{1s}-i\,\frac{\Gamma_{1s}}{2}= -\frac{2\alpha^{3}\mu^{2}c^{4}}{\hbar c}\,a_{\pi D}\left[1-\frac{2\alpha\mu c^2}{\hbar c}\,(\ln\,\alpha -1)\cdot a_{\pi D}\right].\label{eq:Rusetski}
\end{equation}
The first term corresponds to the classical Deser formula\,\cite{Des54} yielding the scattering length in leading 
order\,\cite{Tru61}. The term in brackets corrects for the fact that $a_{\pi D}$ is determined from a Coulomb 
bound state\,\cite{Lyu00,Mei06,Gas08}. In equations (\ref{eq:Rusetski}) and (\ref{eq:true_im}), $\alpha $ denotes the fine 
structure constant. 

The imaginary part of $a_{\pi D}$ is given by 
\begin{eqnarray}
\Im\,a_{\pi D}&=&\frac{\hbar c}{2\alpha^{3}\mu^{2}c^{4}}\cdot\frac{\Gamma_{1s}/2}{1-\frac{2\alpha\mu c^2}{\hbar c}\,(\ln\,\alpha -1)\cdot 2\,\Re\,a_{\pi D}}\nonumber\\
              &=& 0.010642\,\,m^{-1}_{\pi}\,\mathrm{eV}^{-1}\cdot 1.004 \cdot\left(\Gamma_{1s}/2\right)\,.\label{eq:true_im}
\end{eqnarray}
The factor 1.004 stands for the---in this case---small bound-state correction. Hence, it is sufficient 
to insert the leading order result for the real part of the scattering length in equation\,(\ref{eq:true_im}). The 
value $\Re\, a_{\pi D}=(26.3\pm\,0.6)\cdot 10^{-3}\,m^{-1}_{\pi}$ is taken from a previous experiment\,\cite{Hau98}.

Detailed balance states that\,\cite{Bru51}\\
\begin{eqnarray}
\tilde{\sigma}_{\pi^{+}d\rightarrow\,pp}& = &\frac{2}{3}\cdot \left(\frac{p^{*}_{p}}{p^{*}_{\pi}}\right)^2\cdot \,\tilde{\sigma}_{pp\rightarrow~\pi^{+}d}\,,
\end{eqnarray}
where $p^{*}_{p}$ and $p^{*}_{\pi}$ are the final state momenta of proton and pion in the center-of-mass (CMS). 
Assuming charge symmetry, for the transition matrix elements 
$\mid \tilde{M}_{\pi^-d\rightarrow nn}\mid\,=\,\mid \tilde{M}_{\pi^+d\rightarrow pp}\mid$ holds. A  
small difference in the transition rate, 
$\tilde{\sigma}_{\pi^-d\rightarrow nn}/\tilde{\sigma}_{\pi^+d\rightarrow pp}=p^{*}_{n}/p^{*}_{p}=0.982$,
must be taken into account because of the slightly larger phase space for $\pi^+d\rightarrow pp$ with $p^{*}_{\,n,p}$ 
being the nucleon CMS momenta. 

In principle, both electromagnetic and hadronic isospin-breaking effects must be 
considered in view of the different quark contents in the final states of the processes $\pi^-d\rightarrow nn$ 
and $\pi^+d\rightarrow pp$. Their magnitude, however, is assumed to be at most a few per cent\,\cite{Fil09,Bar09}, 
which is about the precision achieved in this experiment but far below the fluctuations of the pion-production data.
The atomic binding energy of the $\pi^-$D system is neglected.

Combining optical theorem, charge symmetry, detailed balance and inserting the s-wave part from (\ref{eq:pid_pp}), 
the purely hadronic imaginary part of the scattering length $a_{\pi^{-}d\rightarrow nn}$ reads in terms of the 
threshold parameter $\alpha$
\begin{eqnarray}
\Im\,a_{\pi^- d\rightarrow nn}&=&\lim_{p^*_{\pi}\to\,0}~\frac{p^*_{\pi}}{4\pi }\cdot \tilde{\sigma}_{\pi^-d\rightarrow nn}\nonumber\\
             &=&\frac{1}{6\pi}\cdot \frac{p^*_{p}\cdot p^*_{n}}{m_{\pi}}\cdot\alpha\,.\label{eq:Ima_alpha} 
\end{eqnarray}

To relate $\Im\,a_{\pi^- d\rightarrow nn}$ to $\Im\,a_{\pi D}$, a correction for 
final states other than $nn$ must be applied. The measured branching ratios \cite{Hig81,Jos60,Don77} yield 
for the relative strength of true absorption with respect to all other processes 
$S'=nn/(nn\gamma+nne^+e^-+nn\pi^0)=$ $2.76\pm 0.04$. Consequently,   

\begin{eqnarray}
\Im\,a_{\pi D}&=&(1+1/S')\cdot \Im\,a_{\pi^- d\rightarrow nn}\nonumber\\
              &=&(2.48\pm 0.01)\cdot 10^{-5}\,\cdot\,\alpha\,\cdot\,m^{-1}_{\pi}\,\mu\mathrm{b}^{-1}\,\label{eq:Ima_nr}.
\end{eqnarray}

\setlength{\tabcolsep}{1.1mm}
\begin{table}[t]
\begin{center}\vspace{-2.2mm}
\caption{Threshold parameter $\alpha $ derived from the hadronic broadening $\Gamma_{1s}$ in $\pi$D 
         and pion-production and absorption data together with a selection of theoretical approaches.}
\label{table:alpha}
\begin{tabular}{cllclc}
\hline\hline\\[-3mm]
\multicolumn{2}{l}{\it{pionic~deuterium}}     &\multicolumn{3}{c}{$\alpha\,/\,\mu$b}  &\\[0mm]
\hline\\[-3.0mm]
$3p$&$\rightarrow ~~1s$                       & 220 & $\pm$& 45   &\cite{Cha9597} \\[0mm]
$2p$&$\rightarrow ~~1s$                       & 257 & $\pm$& 23   &\cite{Hau98} \\[1mm]
$3p$&$\rightarrow ~~1s$                       & 252 & ${\,+\,\atop\,-\,}$&${\,5\atop\,11}$&{\em this exp.}\\[1mm]
\hline\\[-3mm]
\multicolumn{2}{l}{\it{pion~production/absorption}}      &\multicolumn{3}{c}{$\alpha\,/\,\mu$b}  &\\[0mm]
\hline\\[-3.0mm]
$pp$&$\rightarrow ~~d\pi^+$                   &138 & $\pm$& 15$^{*}$ &\cite{Cra55} \\
$pp$&$\rightarrow ~~d\pi^+$                   &240 & $\pm$& 20$^{*}$ &\cite{Ros67} \\
$pp$&$\rightarrow ~~d\pi^+$                   &180 & $\pm$& 20$^{*}$ &\cite{Ric70} \\
$pp$&$\rightarrow ~~d\pi^+$                   &228 & $\pm$& 46       &\cite{Aeb76} \\
$np$&$\rightarrow ~~d\pi^0$                   &184 & $\pm$& 14       &\cite{Hut91} \\
$d\pi^+$&$\rightarrow ~~pp$                   &174 & $\pm$& 3$^{*}$  &\cite{Rit91} \\
$p_{pol}p$&$\rightarrow ~~d\pi^+$             &208 & $\pm$& 5$^{*}$  &\cite{Hei96} \\
$pp$&$\rightarrow ~~d\pi^+$                   &205 & $\pm$& 9$^{*}$  &\cite{Dro98} \\
\hline\\[-3mm]
\multicolumn{2}{l}{\it{theoretical~approach}} &\multicolumn{3}{c}{$\alpha\,/\,\mu$b}  &\\[0mm]
\hline\\[-3.0mm]
\multicolumn{2}{l}{Watson-Brueckner}          &140 & $\pm$& 50       &\cite{Ros54} \\
\multicolumn{2}{l}{rescattering}              &146 &      &          &\cite{Kol66} \\
\multicolumn{2}{l}{rescattering}              &201 &      &          &\cite{Rei69} \\
\multicolumn{2}{l}{Faddeev (Reid soft core)~~~~~~~~~~~~~~~} &220 & & &\cite{Afn74} \\
\multicolumn{2}{l}{Faddeev (Bryan-Scott)}     &267 &      &          &\cite{Afn74} \\
\multicolumn{2}{l}{heavy meson exchange}      &203 & $\pm$& 21       &\cite{Hor93,Nis96} \\
\multicolumn{2}{l}{$\chi $PT NLO}             &220 & $\pm$& 70       &\cite{Len06} \\
\hline\hline\\[-3mm]
\multicolumn{6}{l}{\it{* experiments reporting statistical uncertainty only}}\\[-6mm]
\end{tabular}
\end{center}
\end{table}

Our $\pi$D experiment was performed at the high-intensity low-energy pion beam $\pi$E5 of the proton 
accelerator at PSI by using the cyclotron trap II and a Johann-type Bragg spectrometer equipped with a 
Si crystal and an array of 6 charge-coupled devices (CCDs) as position-sensitive x-ray detector. The 
set-up for the $\pi$D$(3p-1s)$ measurement is similar to the one used for the $\mu$H$(3p-1s)$ 
transition\,\cite{PSI98,Got08,Cov09} but with restricting the reflecting area of the crystal of 100\,mm 
in diameter to 60\,mm horizontally to keep the Johann broadening small. 

\begin{figure}[t]
\begin{center}
\resizebox{0.4\textwidth}{!}{\includegraphics{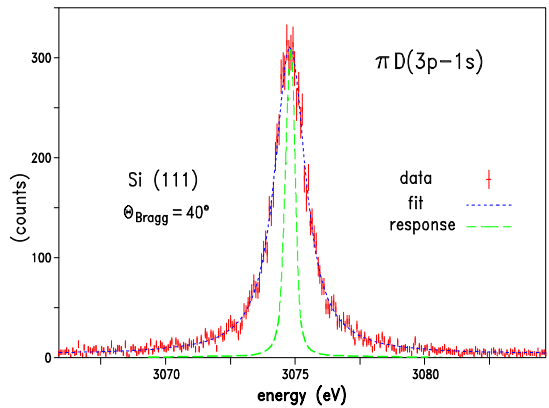}}
\caption{(color online). Sum spectrum of the $\pi$D$(3p-1s)$ transition measured at 10 and 17.5\,bar equivalent density 
         in first order with a silicon (111) Bragg crystal having a bending radius of $R=2982.2\pm 0.3$\,mm. 
         The narrow structure inside the $\pi$D line represents the spectrometer response function.}
\label{figure:piD}
\end{center} 
\end{figure}

The x-ray energy spectrum is obtained by projection of the Bragg reflection onto the axis of 
dispersion\,(Fig.\,\ref{figure:piD}). The granularity of the CCDs having a pixel size of 40\,$\mu$m allows 
an efficient background rejection by means of pattern recognition. Together with a massive concrete shielding, 
x-ray spectra with outstanding peak-to-background ratio are achieved. The stability of the mechanical setup 
was monitored by two inclinometers. Details may be found elsewhere\,\cite{Str09}.

$\pi$D data were taken at equivalent pressures of 3.3, 10, and 17.5\,bar (STP) in order to identify or to exclude 
any x-ray line broadening due to radiative deexcitation from molecular states\,\cite{Taq88,Lin03,Kil04} by 
means of an energy dependence of the $\pi$D$(3p-1s)$ energy. In total, about 1450, 4000, and 4900 $\pi$D events 
were collected corresponding to count rates of 12, 35, and 40 per hour. As in the case of pionic \cite{Got08} 
and muonic hydrogen \cite{Cov09}, no evidence was found for radiative decay after molecule formation within the 
experimental accuracy. The $D_2$ density was adjusted by the temperature of the target gas. 

The $\pi$D line shape is determined by the spectrometer response (i), the natural width of the x-ray transition (ii), 
and Doppler broadening from Coulomb transitions (iii):\\ 
(i) The spectrometer response was determined using the narrow M1 x-ray line from helium-like argon at 3.104\,keV
as outlined in\,\cite{Ana05,Tra07}. When scaled to the $\pi$D$(3p-1s)$ energy (3.075\,keV), it corresponds to a 
resolution of 436$\,\pm$\,3\,meV (FWHM) (Fig.\,\ref{figure:piD}), which is close to the theoretical value of 
403\,meV as calculated for an ideal flat crystal (code XOP\,\cite{San98}).\\ 
(ii) The natural linewidth is practically given by the hadronic broadening $\Gamma_{1s}$. The 3p-level 
width is dominated by radiative decay (28\,$\mu$eV). Nuclear reactions are estimated to be $<$1\,$\mu$eV. 
Likewise, based on calculated transitions rates\,\cite{JeMa02} the induced width due 
to $3p\leftrightarrow 3s$ Stark mixing turns out to be as small as 1\,$\mu$eV. \\
(iii) Coulomb transitions may occur when excited exotic hydrogen atoms penetrate the electron cloud of 
target atoms, and the energy release of the deexcitation step is converted into kinetic energy shared by the 
$\pi$D system and another D atom\,\cite{BF78}. Coulomb deexcitation generates peaks in the 
kinetic energy distribution, which are at 12, 20, 38, and 81\,eV 
for the $\pi$D $\Delta n=1$ transitions (7-6), (6-5), (5-4), and (4-3), respectively. Therefore, 
subsequent x-ray transitions may be Doppler broadened. Acceleration, however, is counteracted by elastic and 
inelastic scattering, which may lead to a continuum below the peak energies or even complete deceleration.

Cascade calculations have been extended to follow the velocity change during the deexcitation 
cascade and, therefore, can provide kinetic energy distributions at the instant of x-ray emission from a 
specific atomic level (extended standard cascade model ESCM\,\cite{JeMa02}). At present, only calculations 
for $\mu$H and $\pi$H are available\,\cite{JeMa02,JPP07,PP07}. 

Therefore, an approach independent of a cascade model was used to extract the Doppler broadening directly 
from the data, which was applied successfully in the neutron time-of-flight and $\mu$H  x-ray ana\-lyses. 
The kinetic energy distribution was modeled by boxes of a few eV width corresponding to  
the peaks generated by the Coulomb transitions. Their number, width, and position are preset but adjustable  
parameters of the analysis code. The hadronic width $\Gamma_{1s}$, the total intensity, and the background 
level are free parameters of the fit, as are the relative intensities, whose sum is normalized to one. 

The line shape was constructed by convoluting the crystal response, imaging properties of the bent crystal, and 
Doppler contributions with the natural linewidth by means of Monte-Carlo ray-tracing. 
Following the experience of the $\mu$H$(3p-1s)$ ana\-lysis\,\cite{Cov09}, one tries to identify consecutively 
individual Doppler contributions starting with one single box only and moving it through the range of possible kinetic 
energies\,\cite{Str09}. A $\chi^2$ analysis using the MINUIT package\,\cite{Jam75} shows the necessity of a 
low-energy contribution. It was found that the upper bound of this box must not exceed 8\,eV, a result obtained 
independently for the spectra taken at 10\,bar and the 17.5\,bar equivalent density. The result for $\Gamma_{1s}$ 
turned out to be insensitive to the upper boundary of the kinetic-energy box for values $\leq 8$\,eV. The 
low-energy component was set to the range $0-2$\,eV in further analysis. 

Searches for any contributions of Coulomb transitions leading to higher energies failed---also for the 
sum of the spectra measured at 10 and 17.5\,bar equivalent density. Tentatively, we used the kinetic energy 
distribution from the ESCM calculation for the $\pi$H$(3p-1s)$ case, after scaling to $\pi$D energies. This 
distribution, where a fraction of 25\% has energies above 15\,eV, is unable to describe the line shape, which 
is in strong disagreement with the findings for muonic\,\cite{Cov09} and pionic hydrogen\,\cite{Sch01,Got08}, 
where sizeable contributions from higher energy Coulomb transitions are mandatory. There is no explanation 
yet for this different behavior in pionic deuterium.

Detailed Monte-Carlo studies have been performed to quantify which amount of high-energy components may be 
missed for the statistics achieved. It was found that a contribution of 25\% can be excluded at the level 
of 99\% for the component around 80\,eV corresponding to the ($4-3$) Coulomb transition. The chance to 
identify a 10\% contribution is about 2/3 corresponding to 1$\sigma$. The value for $\Gamma_{1s}$ itself 
hardly varies with the width of any assumed high-energy box. 

Omitting any high-energy contribution yields the upper limit for $\Gamma_{1s}$. Defining the limit of 
sensitivity  to 10\% -- according to the above-mentioned 1$\sigma$ criterion -- results 
in a lower bound 43\,meV lower than the upper limit and, hence, in an asymmetric systematic error for 
$\Gamma_{1s}$. 

Systematic discrepancies (bias) as arising in maximum likelihood fits\,\cite{Ber02} have been studied according 
to the statistics of the measured spectra. The bias has been determined to be -32$\pm$2\,meV (3.3\,bar data) and 
-2$\pm$2\,meV (sum of 10 and 17.5\,bar). For each set of conditions 400 Monte-Carlo generated spectra were analyzed. 

The weighted average for the hadronic broadening, 
\begin{eqnarray}
\Gamma_{1s}=\left(1171\,{+\,~23\atop-\,~49}\right)\,\,\mathrm{meV}\,,\label{eq:Ga_final}
\end{eqnarray}
is in good agreement with previous measurements which found $\Gamma_{1s}=1020\,\pm$\,210\,\cite{Cha9597} 
and 1194\,$\pm$\,105 meV\,\cite{Hau98}.

This result leads to an imaginary part (\ref{eq:true_im})   
\begin{eqnarray}
\Im\, a_{\pi D} &=& \left(6.26 {+\,0.12\atop -\,0.26}\right)\cdot 10^{-3}\,m^{-1}_{\pi}\,.\label{eq:Tcorr}
\end{eqnarray}
The corresponding value for $\alpha$ is given in Table\,\ref{table:alpha}.

In summary, the $\pi$D$(3p-1s)$ x-ray transition in pionic deuterium has been studied to determine the 
strong-interaction broadening of the 1s state and from that the threshold pion-production strength $\alpha$. 
The accuracy of 4.2\% achieved reaches the expected $(5-10)$\% uncertainty of forthcoming NNLO 
$\chi$PT calculations. It is noteworthy that at the 10\% level no components from high-energetic Coulomb 
transitions could be identified.
\begin{acknowledgments}
We thank N.\,Dolfus, L.\,Stohwasser, and K.-P.\,Wieder for their technical assistance and C.\,Hanhart and 
A.\,Rusetsky for continuous exchange on theoretical progress for the $\pi$D system. The Bragg crystal was 
manufactured by Carl Zeiss AG, Oberkochen, Germany. 
Partial funding and travel support was granted by FCT (Lisbon) and FEDER (PhD grant SFRH/BD/18979/ 2004 and project 
PTDC/FIS/102110/2008) and the Germaine de Sta$\ddot{e}$l exchange program. LKB is Unit\'{e} Mixte de Recherche 
du CNRS, de l'\'{E}cole Normale Sup\'{e}rieure et de UPMC n$^{\circ}$ C8552. This work is part of the PhD 
thesis of one of us (Th.\,S., Univ. of Cologne, 2009). 
\end{acknowledgments}

\end{document}